\documentclass[11pt,eucal]{article}
\usepackage{amsmath}
\usepackage{amssymb}
\usepackage{epic}
\usepackage{eepic}
\usepackage{times}
\usepackage{graphics}

\begin{document}

\title{Algorithmic Cooling and Scalable \\ NMR Quantum
 Computers\thanks{This work was sponsored in part by the Defense 
Advanced Research Projects
Agency (DARPA) project MDA972--99--1--0017 [note that the content of this
paper does not necessarily reflect the position or the policy of the
government, and no official endorsement should be inferred], and in part
by the U.S. Army Research Office/DARPA under contract/grant number
DAAD19--00--1--0172.
\newline\mbox{\hspace{4mm}}
The research of Farrokh Vatan was performed partly at UCLA and supported 
by the above mentioned grants
and was performed partly at the Jet Propulsion Laboratory
(JPL), California Institute of Technology, under contract with National 
Aeronautics and Space Administration (NASA). The Revolutionary computing
Technologies Program of the JPL's Center for Integrated Space 
Microsystems (CISM) supported his work.}}
\author{P. Oscar Boykin$^1$, Tal Mor$^{1,2 \, *}$,
Vwani Roychowdhury$^1$, \\ Farrokh Vatan$^{1,3}$, and Rutger Vrijen$^4$
\\ \small 1. Electrical Engineering Department,  UCLA,
Los Angeles, CA 90095, USA.
\\ \small 2. Electrical Engineering Department, College of Judea and Samaria, 
Ariel, Israel.
\\ \small 3. Jet Propulsion Laboratory, California Institute of Technology,
             4800 Oak Grove Drive Pasadena, CA 91109 
\\ \small 4. Sun Microsystems Laboratories, Mountain View, CA, USA             
\\ \small * To whom correspondence should be addressed. 
Email: talmo@cs.technion.ac.il}

\date{}

\maketitle

\begin{abstract}

We present here {\em algorithmic cooling (via polarization-heat-bath)}---a
powerful method for obtaining a large number of highly polarized spins 
in liquid 
nuclear-spin systems at finite temperature. Given that spin-half states 
represent (quantum) bits,
algorithmic cooling cleans dirty bits {\em beyond} the Shannon's bound 
on data compression, 
by employing a set of rapidly thermal-relaxing bits. 
Such auxiliary bits could be implemented using spins that  rapidly get 
into thermal equilibrium with the environment, e.g., electron spins.

Cooling spins to a very low temperature
{\em without} cooling the environment 
could lead to a breakthrough in nuclear magnetic resonance
experiments, and 
our ``spin-refrigerating'' method suggests that this is possible.

The scaling of NMR ensemble computers is probably the main obstacle to 
building useful quantum computing devices, and 
our spin-refrigerating method suggests that this problem can be resolved.

\end{abstract}

\section{Introduction}

Ensemble computing is based on a model comprised of a macroscopic
number of computers, where the same set of operations is performed
simultaneously
on all the computers.  
The concept of ensemble computing became very important recently,
due to the fact that NMR quantum computers~\cite{NMR-exp}
perform ensemble computing. NMR quantum computing has already
succeeded in performing complex operations involving up to 7-8 qubits
(quantum bits), and
therefore, NMR quantum computers 
are currently the most successful quantum computing devices.

In NMR quantum computing
each computer is represented by a single molecule, and the qubits of the
computer are represented by the nuclear spins embedded in a single molecule. 
A macroscopic number of 
identical molecules is available in a bulk system,
and these molecules act as many
computers performing the same
computation in parallel.
To perform a desired computation, the same sequence of
external pulses is applied to all the molecules/computers. 
Finally, a measurement of the state of a single qubit is performed 
by averaging over all computers/molecules to read out the
output on a particular bit on all computers. Due to the use of a macroscopic
number of molecules, the output is a noticeable magnetic signal. 
It has been shown that almost all known quantum algorithms designed for
the usual single-computer model, can be adapted to be implemented on ensemble
computers~\cite{our-bulk}, and in particular,
these ensemble computers can perform fast factorization of
large numbers~\cite{Shor} and fast data-base search~\cite{Grover}.

Unfortunately, the wide-spread belief is that even though ensemble
quantum computation is a powerful scheme for demonstrating fundamental quantum
phenomena, it is not scalable (see for 
instance~\cite{Warren,DiVince,popescu-et-al}). 
In particular, in the current approaches to
ensemble computing, identifying the state of the computer requires sensing
signals with signal-to-noise ratios that are exponentially small in $n$, the
number of qubits in the system. We refer to this well-known problem as the
{\em scaling problem}.
The origin of the scaling problem is explained in the following.

The initial state of each
qubit, when averaged over all computers (a macroscopic number),
is highly mixed, with only a small bias towards the zero state.
At thermal equilibrium the state is
\begin{equation} \label{initial-state}
  \rho_{\epsilon_0} = 
  \begin{pmatrix}
               (1+\epsilon_0)/2 & 0 \\
               0                & (1-\epsilon_0)/2
  \end{pmatrix}, 
\end{equation}
where the initial bias, $\epsilon_0$, 
is mainly determined by the magnetic field and 
the temperature, but also depends on the structure and the 
electronic configurations of the 
molecule. 
For an ideal system, one has $\epsilon_0=\epsilon_{\rm perfect} = 1$ leading to 
$\rho_{\epsilon_{\rm perfect}}= |0\rangle \langle0| =\begin{pmatrix}
  1 & 0 \\
 0  & 0
 \end{pmatrix}$,
meaning that the state is $|0\rangle $ with probability one, and it is
$|1\rangle$ with probability zero.
For a totally mixed system, $\epsilon_0=0$, hence the probabilities of 
$|0\rangle$ and $|1\rangle$ are both equal to half.
We also 
define $\delta_0 = (1 - \epsilon_0)/2$ to be 
the initial error probability.
Typically, $\epsilon_0$ is around $10^{-6}$ for the
liquid NMR systems in use \cite{NMR-exp}, and can probably be improved
(increased) a great deal in the near future. Especially promising
directions are the use of liquid crystal NMR for quantum
computing~\cite{Chuang99}, and the use of a SWAP operation for the nuclear
spin and the electron spin known as ENDOR technique~\cite{kurreck}.     
The state of an $n$-qubit system in the ideal case is 
$\rho^{\{n\}}_{ideal} = |0_n\rangle \langle 0_n|$ with 
$|0_n\rangle = |0\rangle \otimes |0\rangle \otimes \ldots \otimes |0\rangle $
(a tensor product of $n$ single qubit states).
In general, the initial state of an $n$-qubit liquid NMR system 
can be represented as a tensor product of states of the individual qubits:
\begin{equation} \label{rho-n-init}
\rho_{\mathrm{init}}^{\{n\}} = 
\rho_{\epsilon_0} \otimes \rho_{\epsilon_0} \otimes
\cdots \otimes \rho_{\epsilon_0} .
\end{equation}
This state can also be written as
$\sum_{i=0}^{2^n - 1} P_i |i\rangle\langle i|$,  
a mixture of all states $|i\rangle$---the basis vectors of the system,
and $i$ (for $n$ qubits) is a $n$-bit binary string.
E.g., for two qubits, $P_{00}=(1+\epsilon_0)^2 / 4$.
In fact, the initial bias is not the 
same on each qubit~\cite{diff-bias},
but as long as the differences between these biases are 
small we can ignore this fact in
our analysis. The analysis we do later on is correct 
if we replace all these slightly different initial biases by their minimum 
value and call this value $\epsilon_0$.

Currently, researchers use the so-called ``pseudo pure state (PPS)'' technique 
to perform computations with such highly mixed initial states. 
In this technique, the initial mixed density
matrix  is transformed to a state 
\begin{equation} \label{pps}
\rho_{\rm PPS}^{\{n\}}\ \equiv \ (1-p) {\cal I} + p |\psi\rangle\langle \psi|,
\end{equation}
which is a mixture
of the totally-mixed state
${\cal I}=\frac{1}{2^n}I_{2^n}$
(with $I_{2^n}$, the identity matrix of order $2^n$),
and a pure state $|\psi\rangle$ initially 
set to be $|0_n\rangle$ in our case.
Such a state is called a pseudo-pure state. Unitary operations then leave
the totally mixed state unchanged, but do affect the pure state part, to perform
the desired computation via entanglement of the pure part
(which we refer to as ``pseudo-entanglement'').  
Finally, the state of the ensemble-computer is 
measured. If the probability $p$ of the pure state is not too small, then
the pure part of the state yields the {\em expectation value} for each qubit,
an outcome which is sufficient for performing quantum computing which is
as powerful as the standard (non-ensemble) quantum computing~\cite{our-bulk}.
Unfortunately,  in all the existing PPS methods 
\begin{equation} p = \displaystyle \frac{(1+\epsilon_0)^n - 1}{2^n - 1}
 < 2 \left(\frac{1+\epsilon_0}{2}\right)^n , \end{equation}
and hence, $p$ scales exponentially badly with $n$ 
(the number of computation qubits),
leading to an exponentially small signal-to-noise
ratio. As a result, an exponential number of computers 
(molecules) are required in order to 
read the signal.
With $\epsilon_0$ in the range $10^{-6}-10^{-1}$ one might still hope to 
obtain a 20-qubit 
computer, since then $p$ (approximately $10^{-5}-10^{-6}$) 
can still lead to an observed signal
when an Avogadro number of computers are used. 
But one cannot hope to go  beyond
a 50-qubit computer,
since then, $p$ is approximately $10^{-13}-10^{-15}$, which is smaller than 
the standard deviation
in reading the result 
(and, even with perfect devices, the signal cannot be read).

The exponential advantage of quantum computers 
over classical ones~\cite{Shor} is   
totally lost in these NMR computing devices
since an exponential number of molecules/computers is 
required for the computation,
and therefore the scaling problem must be resolved in order to achieve 
any useful NMR quantum computing. This scaling problem (plus the assumption  
that quantum computing requires entanglement, 
and cannot rely on pseudo-entanglement) 
has led several researchers to suggest that the
current NMR quantum computers are no more than classical 
simulators of quantum computers~\cite{popescu-et-al}.
Actually, the important contribution of ~\cite{popescu-et-al} 
is the result that in some
neighborhood of the totally-mixed state, all states are 
separable; hence, some pseudo-entanglement
states contain no entanglement. But this work does 
not prove (and does not claim to prove) 
that current NMR quantum computers do not perform quantum computation.
We, in contrast, conjecture that the PPS technique and
the work of~\cite{popescu-et-al} form the 
first step in proving that 
quantum computing without entanglement {\em is possible}.

The first important step in resolving the scaling problem is to understand that 
the scaling problem is not an inherent 
characteristic of ensemble computers but is an {\em artifact}
of  the existing PPS methods. 
In fact, the original highly mixed state contains a great deal 
of information, and this 
can be seen by rotating each qubit separately and finally measuring the qubits.
However, the existing methods of  transforming the highly mixed state into the PPS cause the 
scaling problem, by losing information {\em on purpose}. 
Furthermore, it is important to mention that for any
$n$, there is a range of bias, $\epsilon$, not close to zero, 
where the currently existing methods
for creating PPS work just fine. 
In order to be in that range, the state of each qubit must be almost pure:
$\epsilon = 1 - 2 \delta$, where $\delta$, 
the error probability, satisfies $\delta \ll 1$ (actually $\delta \approx 0.2$ is already useful).
Then $p$ scales well: 
\begin{equation} p = \displaystyle \frac{(1+\epsilon)^n - 1}{2^n - 1}
 \approx   (1 - \delta)^n .
\label{scale-well}
\end{equation}
As long as $\delta \approx O(1/n)$, the probability $p$ is sufficiently large
for all practical purposes, thus
much larger $n$'s can
still be used. 
Furthermore, {\em any} $n$ can be used if one can control $\delta$ as a function of $n$.
The PPS technique, the loss of information,
and the scaling problem are
described in more detail in appendix~\ref{PPS-tech}.

Instead of converting the initial state (\ref{rho-n-init}) to a PPS
(\ref{pps}), we perform a
``purification'' transformation that takes a subset, $m$ (with $m\leq n$), of
the qubits to a final state of the form
\begin{equation}
\rho_{\mathrm{final}}^{\{m\}}=\rho_{\epsilon_{\mathrm{des}}}\otimes
\rho_{\epsilon_{\mathrm{des}}}\otimes
\cdots\otimes\rho_{\epsilon_{\mathrm{des}}},
\label{rho-f}
\end{equation}
where $\epsilon_{\mathrm{des}}$ is some desired bias close enough to 1. This
state with a higher bias can then be transformed into 
a scalable PPS, $\rho_{\rm PPS}^{\{m\}}$. 
For example, {\em we shall demonstrate how to achieve (via algorithmic cooling)}
  $\delta \approx 0.2$, which allows $n$ in the 
  range 20-50 qubits, and $\delta \approx 0.04$,
which allows $n$ in the range of 50-200 qubits.

In this paper we present a purification process which uses concepts from 
information theory (data compression) and from thermodynamics
(heat bath, thermal relaxation), and which resolves the scaling problem.
Our ``information-theoretic'' purification is totally classical, hence the 
density matrices are treated as classical probability distributions, 
and no explicit quantum effects are taken into consideration.
In earlier work, Schulman and Vazirani~\cite{shul_vaz} already demonstrated novel
compression-based (and not PPS-based) alternative NMR computing,
which does not suffer from the scaling problem. Their scheme is based on information
theoretic tools, and it leads to Eq.~(\ref{rho-f}).
However, the Shannon bound on the purification ability
prevents purifying any reasonable fraction of bits for small values of 
$\epsilon_0$: 
$m\approx \frac{\epsilon_0^2}{2\ln 2} n$ (see Section~\ref{sec-bcs}),
meaning that thousands 
of bits are required in order to get
one or a few purified bits (with a reasonable probability of success).
More explicitly, any entropy-preserving 
purification scheme cannot currently be useful for NMR computation.

We present here the first cooling scheme that goes beyond the Shannon bound,
an {\em algorithmic cooling via polarization-heat-bath},
or in short, {\em algorithmic cooling}.
This cooling scheme, presented in Section~\ref{sec-cool},
purifies a large fraction of the bits initially set in 
a highly mixed state, and hence resolves the scaling problem. 
Algorithmic cooling  can bypass the Shannon bound
since it does not preserve entropy of the system,
but removes entropy into a heat bath at a temperature $\beta_0$.
In order to pump entropy into the polarization heat bath, algorithmic cooling 
demands the existence and the mutual processing of two types of 
qubits~\cite{bits-qubits}:
computation bits and
bits which Rapidly Reach Thermal Relaxation (RRTR bits). 
The computation bits are assumed to have a very long relaxation 
time, ${\cal T}_{\rm comput-bits}$,
and they are used for the computation, 
and the RRTR bits are assumed to have a much shorter 
relaxation time, ${\cal T}_{\rm RRTR}$,
hence they rapidly get into thermal equilibrium with the environment
(a heat bath) at a temperature of $\beta_0$.
Since the RRTR bits are defined via their 
spin (to be 0 or 1), the heat bath is actually a 
spin-polarization heat bath. In our algorithmic cooling, 
a standard compression is performed on the computation bits, 
purifying (cooling) some while concentrating the entropy (heating) the others, 
to heat them above $\beta_0$.
Then the hotter bits are replaced with the RRTR bits, which are at the 
heat-bath temperature $\beta_0$, resulting in an overall cooling of the system.
Repeating the process many times via a recursive algorithm, any final close-to-zero 
``temperature'' (that is, any final bias) can in principle be achieved

Algorithmic cooling provides a 
new challenge for the experimentalists, since such processing of two types of quantum bits
(two different spin systems) is highly nontrivial.
The currently existing experimental technologies, and the new
``experimental challenge'' of combining them in order to perform
algorithmic cooling, are explained further in Section~\ref{sec-exp}.
Conclusions and some open questions for further research are provided 
in Section~\ref{conc}.

\section{Information Theory, the Basic Compression Subroutine  
and Purification Levels}
\label{sec-bcs}

\subsection{Shannon's bound}

Let us briefly describe the
purification problem from an information theoretic perspective. 
There exists a straightforward correspondence between the initial state
of our $n$-qubit system, 
and a probability distribution of all $n$-bit binary
strings, where the probability of each string $i$ is given by the 
term $P_i$, the probability of the state $|i\rangle$ in the mixed state
$\rho_{\mathrm{init}}^n$ described by Eq.(\ref{rho-n-init}).
A~loss-less compression of a random
binary string which is distributed as stated above has been well studied. 
In an optimal compression scheme, all the randomness (and hence,
the entropy) of the bit string is transferred to $n-m$ bits, while 
with extremely high probability leaving
$m$ bits in a known deterministic state, say the string $0$.
The entropy $H$ of the entire system is 
$H({\rm system}) = n H({\mathrm single-bit}) = n H(1/2 + \epsilon_0/2)$ with 
$H(P) \equiv - P \log_2 P - (1-P) \log_2 (1-P)$ measured in bits. 
Any loss-less compression scheme preserves 
the entropy $H$ of the entire system, 
hence, one can apply Shannon's source coding
bound on  $m$ to get $m \le n[1 - H(1/2+\epsilon_0/2)]$. 
Simple leading-order calculation shows that $m$ is bounded by (approximately) 
$\frac{\epsilon_0^2}{2\ln 2} n$ for small values of the initial bias
$\epsilon_0$, and in a practical compression scenario
this can be achieved if a large enough string
(large enough $n$) is used.
Schulman and Vazirani~\cite{shul_vaz} were the first to use information
theoretic tools for solving the scaling problem, 
and they also demonstrated how to
get very close to the Shannon bound, once $n$ is very large.
We consider here a bias of $0.01$ and a bias of $0.1$,
and with these numbers, the Schulman-Vazirani compression cannot
be useful in practice, 
and cannot help in achieving NMR computing with more than 
20 qubits in the foreseeable future.
In fact, any entropy-preserving purification scheme cannot be useful for NMR
computation in the near future. 

We suggest here an entropy-nonpreserving
purification.
Our purification, algorithmic cooling, has some common properties 
with the entropy-preserving purification,
such as the basic compression subroutine
and the purification levels. These are therefore described in the following.

\subsection{Basic Compression Subroutine and Purification Levels}

The Basic Compression Subroutine
(BCS) is the simplest purification procedure used to 
convert a mixture with a particular
bias $\epsilon_j$,  to one with a higher bias $\epsilon_{j+1}$ but fewer bits. 
We take pairs of bits and check if they are the
same or different. One bit (the ``supervisor'') 
retains the information of whether or not they were the same. 
If they were the same, then we keep the other bit (the ``adjusted'' bit)
and we say it is purified.
This way we increase the bias or {\em push the bits to a higher
purification level}. 
To realize this operation we use a Controlled-NOT (CNOT) transformation
on a control bit ($c$) and a target bit ($t$):
$0_c 0_t \rightarrow 0_c 0_t$,
$0_c 1_t \rightarrow 0_c 1_t$,
$1_c 0_t \rightarrow 1_c 1_t$,
$1_c 1_t \rightarrow 1_c 0_t$.
After the transformation, the target bit holds the information
regarding the 
identity of the initial states of the two bits, hence it is the supervisor bit.
If the target bit is $0$ after the CNOT operation
between a pair of bits, then the pair had the same initial value and the control
bit of the CNOT (the adjusted bit) is retained since it is {\em purified},
otherwise they were different and the adjusted bit
is thrown away since it got dirtier. 
In both cases, the supervisor bit has a reduced bias (increased entropy),
hence it is thrown away.
However, before being thrown away, the
supervisor bit is used as a control bit for a SWAP operation:
if it has the value ``0'', then it SWAPs the
corresponding adjusted bit at the
head of the array (say to the left), and if it is ``1'' it leaves the 
corresponding adjusted bit at its current place. In either cases the
supervisor bit is then SWAPped to the right of the array.
[Note that we use here  
a hybrid of English
and symbol languages to describe an operation, such as SWAP or CUT.] 
As a result, at the end of the BCS all purified bits are at the first locations
at the left side of the array, the dirty adjusted bits are at the center,
and the supervisor bits 
are at the right side of the array. Thus the dirty adjusted bits and the 
supervisor bits can be thrown away (or just ignored).

Starting a particular BCS on an even number $n_j$ of bits with 
a bias  $\epsilon_j$, at the end of the BCS there are 
(on average) $n_{j+1}$ purified
bits with a new bias $\epsilon_{j+1}$. 
The new length and new bias are calculated as follows.
The probability of an adjusted bit
being $|0\rangle$ in the purified mixture, i.e., $(1+\epsilon_{j+1})/2$,
is obtained by a direct application of Bayes' law and is given by:
$  \left[1+\epsilon_{j+1} \right]/2 = \frac{P_{00}}{P_{00} + P_{11}} 
=\frac{(1+\epsilon_j)^2 / 4}{(1+\epsilon_j^2)/2}=
   \left[1+\frac{2\epsilon_j}{1+\epsilon_j^2}\right]/2 $,     
where the $P_i$ are defined for a 2-bit string, so that 
$P_{00}=\frac{1+\epsilon_j}{2}\frac{1+\epsilon_j}{2}$,
and $P_{11} =\frac{1-\epsilon_j}{2}\frac{1-\epsilon_j}{2}$.
The new bias is
\begin{equation}
 \epsilon_{j+1}=\frac{2\epsilon_j}{1+\epsilon_j^2} .
\label{eq-epsilon}
\end{equation}

The number of purified bits, ${\cal L}_{j+1}$, with the new 
bias $\epsilon_{j+1}$
is different on each molecule.
Since, for each pair,
one member is kept with
probability $P_{00}+P_{11}=(1+\epsilon_j^2)/2$, 
and the other member is thrown away,
the expected value of the
length of the purified string is 
\begin{equation}
n_{j+1} =  \langle {\cal L}_{j+1} \rangle = \frac{1+\epsilon_j^2}{4}n_j .
 \label{frac1}
\end{equation}
Note that $n_{j+1} = (\epsilon_j/2\epsilon_{j+1}) n_j$.

The number of steps in one such BCS is calculated as follows.
There are $n_j / 2$ pairs. For each pair one CNOT operation is performed. 
Then, at most $\frac{3 n_j}{2} - 3$ (that is, less than $2n_j - 1$)
operations of controlled-SWAPs and SWAPs are performed to
conditionally put the 
adjusted bit in the first location at the left of the array,
or leave it in its current location:
first, a controlled-SWAP is performed with the supervisor bit as a control, 
the adjusted bit and the bit to its left as the target.
Then the SWAP operation is performed on the supervisor bit and the 
bit which is one location
to its left. 
Then a controlled-SWAP is again performed
to conditionally swap the two bits at the left of the supervisor bit,
and again the supervisor bit is SWAPped
one location to the left.
These SWAP and controlled-SWAP operations
are then repeated until the adjusted bit is 
conditionally SWAPped all the way to the 
first location of the array 
[the supervisor bit is at the third location in the array
when this final controlled-SWAP is performed].
Finally, the supervisor bit is SWAPped till it 
reaches the previously used supervisor bit.  At 
the end of these operations
all used supervisors are at the right of the array, 
all purified adjusted bits are at 
the left of the array, and all the adjusted bits which got dirtier are
to the right of the purified adjusted bits.
Considering controlled-SWAP, SWAP, 
and CNOT as being a single operation each 
(hence one time step each), we obtain a total of 
\begin{equation} \label{time-steps-BCS}
T_{\rm BCS} < (2 n_j)(n_j/2) = {n_j}^2 
\end{equation}
time steps for a single BCS operation. 
Actually, even if each controlled-SWAP is considered as two time steps 
this bound still holds, once a more tight bound is calculated.

A full compression scheme can be built by repeating the BCS several
times, such that the first application, ${\cal B}_{\{0 \rightarrow 1\}}$,
 purifies the bits from 
$\epsilon_0$ to $\epsilon_1$, 
and the second purification, ${\cal B}_{\{1 \rightarrow 2\}}$, 
acts only on bits that were already purified to
$\epsilon_1$, and purifies them further to $\epsilon_2$.
The $j^{\rm th}$ application, ${\cal B}_{\{(j-1) \rightarrow j\}}$, 
purifies bits from 
$\epsilon_{j-1}$ to $\epsilon_j$. 
Let the total number of BCS steps be
$j_{\rm final} \equiv j_f$ and let the final bias achieved after $j_f$ 
applications of the compression be
$\epsilon_{j_{\rm final}} \equiv \epsilon_f$.
By iterating equation (\ref{eq-epsilon}) we calculate directly
$\epsilon_{j_{\mathrm{final}}}$ 
when starting with $\epsilon_0 = 0.01$ or $\epsilon_0 = 0.1$,
and after $j_{\mathrm{final}}$ application of BCS.
Then, using $\delta_{\rm final} = (1 - \epsilon_{\rm final})/2$,
and Eq.(\ref{scale-well}) with $m$ purified bits, we estimate the 
number of bits $m$ for which a scalable PPS technique can be obtained. 
The results are summarized in Table 1. 
The first interesting cases within the table are
$\epsilon_0 = 0.01$; $j_f=6$, or $\epsilon_0 = 0.1$; $j_f=3$, 
allowing up to $m=50$ bits. 
We refer to these possibilities as a short term
goal.
As a long term goal, up to 
200 bits can be obtained with 
$\epsilon_0 = 0.01$; $j_f=7$, or  $\epsilon_0 = 0.1$; $j_f=4$. 
We consider cases in which the probability $p$ of the pure state
is less than $10^{12}$ as unfeasible.

When only the BCS is performed, the resulting 
average final length of the string 
is $m = n_{j_f} = (\epsilon_{j_f-1}/2\epsilon_{j_f}) n_{j_f-1} =
 (\epsilon_{j_f} / 2^{j_f} \epsilon_0) n_0$,
so that the initial required number of bits, $n_0$, is huge,
but better compression schemes can be designed~\cite{shul_vaz},
which approach the Shannon's bound.
However, for our purpose, which is to achieve 
a ``cooling via polarization-heat-bath''
algorithm,
this simplest compression scheme is sufficient.

\section{Algorithmic Cooling via Polarization-Heat-Bath}
\label{sec-cool}

\subsection{Going Beyond Shannon's Bound}

In order to go {\em beyond  Shannon's bound } 
we assume that we have a thermal bath of partially polarized bits 
with a bias $\epsilon_0$.
More adequate to the physical system, we assume that
we have rapidly-reaching-thermal-relaxation (RRTR) bits.
These bits, by interaction with the environment at some constant 
temperature $\beta_0$,
rapidly return
to the fixed initial distribution with bias of $\epsilon_0$ 
(a reset operation).  
Hence, the environment acts as a polarization heat bath.

In one application of the BCS on bits at a bias of $\epsilon_j$,  
some fraction $f$ (satisfying $1/4 \le f \le 1/2$)
is purified to the next level, $\epsilon_{j+1}$
while the other bits have increased entropy. 
The supervisor bits are left with a reduced bias of 
${\epsilon_j}^2$, and the adjusted bits which failed to be purified
are changed to a bias  $\epsilon = 0$,
that is, they now remain with full entropy.

To make use of the heat bath for removing entropy, we swap a {\em
dirtier}
bit with an RRTR bit at bias $\epsilon_0$, and do not use 
this RRTR bit until it thermalizes back to  $\epsilon_0$.
We refer to this operation as a single ``cooling'' 
operation~\cite{cooling}.
In a nearest-neighbor gate array model, which is  
the appropriate model for NMR quantum computing,
we can much improve the efficiency of the cooling by assuming that each 
computation bit has
an RRTR bit as its neighbor 
(imagine a ladder built of a line of computation bits 
and a line of RRTR bits). 
Then $k$ cooling operations can be done in a single time step 
by replacing $k$ dirty bits with $k$ RRTR bits in parallel.

By applying many BCS steps and cooling steps in a recursive way, spins can be
refrigerated to any temperature, via algorithmic cooling.

\subsection{Cooling Algorithm}\label{cool-algo}

For the sake of simplicity, 
we design an algorithm whereby BCS steps are always applied to 
blocks of exactly $m$ bits (thus, $m$ is some pre-chosen even constant),
and which finally provides $m$ bits at a bias $\epsilon_{j_f}$.
Any BCS step is applied onto an array of $m$ bits at a bias $\epsilon_j$,
all purified bits are pushed 
to the head of the array (say, to the left), all 
supervisor bits are swapped to 
the back of the array (say, to the right), and all
unpurified adjusted bits (which actually became much dirtier) are 
kept in their place.
After one such BCS step, the $m/2$ bits at the right have bias of
${\epsilon_j}^2$, the purified bits at the left have a bias $\epsilon_{j+1}$,
and to their right there are bits with a bias zero. Note that the 
boundary between
the purified adjusted bits and the dirtier adjusted bits is known only by its
expected value $\langle {\cal L}_{j+1} \rangle = \frac{1+ \epsilon_j^2}{4}\ m$.
By repeating this set of operations $\ell$ times (as explained in 
the following paragraphs),
with $\ell \ge 4$, 
an expected value $\langle {\cal L}^\ell_{j+1} \rangle =
 \frac{\ell(1+ \epsilon_j^2)}{4}\ m$
of bits is obtained, from which the first $m$ bits are defined as the output bits
with $\epsilon_{j+1}$, and the rest are ignored. 
If an additional purification is now performed,
only these first $m$ bits are considered as the input for that purification.
We refer to $\ell$ as the ``cooling depth'' of the cooling 
algorithm~\cite{ell-j}.

The algorithm is written recursively with 
purification-steps $M_j$, where the $j^{\mathrm{th}}$ purification step 
corresponds to purifying an initial array of $N_j$ bits into a  
set of $m$ bits at a bias level of $\epsilon_j$, via 
repeated compression/cooling  
operations described as follows: 
In the purification step $M_0$ we wish to obtain $m$ bits with a bias
$\epsilon_0$. In order to achieve this
we SWAP $m$ bits with $m$ RRTR bits, which results in 
$m$ cooling operations performed in parallel.
The number of bits required for $M_0$ is $N_0 = m$. 
In one purification step $M_{j+1}$ (with $j \ge 0$) we wish to 
obtain $m$ bits with a bias $\epsilon_{j+1}$.
In order to achieve this goal
we apply $\ell$ purification steps $M_{j}$, each followed by 
a BCS applied to exactly $m$ bits at a bias $\epsilon_{j}$.
First, $M_j$ is applied onto $N_j$ bits, yielding an output of $m$ bits at
a bias $\epsilon_j$. A BCS is then applied onto these bits, yielding
a string of expected length 
$\langle {\cal L}_{j+1}^1 \rangle = \frac{1+ \epsilon_j^2}{4} \  m$
bits purified to a bias $\epsilon_{j+1}$ and pushed all the way to the left.
At the end of that BCS all the $m/2$ supervisor bits are located at positions
$m/2 + 1$ until $m$.
Then $M_j$ is applied again onto an array of $N_j$ bits, starting at position
$m/2 + 1$.
This time all BCS operations {\em within} this second 
application
of $M_j$ push the bits to the relative first  
location of that $M_j$ array which is 
the location $m/2 +1$ of the entire string.
[In the case of $j=0$, of course, 
there are no BCS operations within $M_0$.]
At the end of that second $M_j$ application, a BCS is applied to 
$m$ bits at a bias
$\epsilon_j$ (at locations $m/2 + 1$ till $m/2 + m$),
purifying them to $\epsilon_{j+1}$. The purified bits are pushed 
{\em all the way} to the left, leading to a string of expected length 
$\langle {\cal L}_{j+1}^2 \rangle =2 \frac{1+ \epsilon_j^2}{4} \  m$.
At the end of that BCS all the $m/2$ supervisor bits are located at positions
$m + 1$ till $3m/2$.
Then $M_j$ is again applied onto an array of $N_j$ bits, starting at position
$m + 1$.
All BCS operations {\em within} this third  
application of $M_j$ push the bits to the relative first  
location of that $M_j$ array (the location $m +1$ of the entire string).
At the end of that third $M_j$ application, a BCS is 
applied to $m$ bits at a bias
$\epsilon_j$ (at locations $m + 1$ till $m + m$),
purifying them to $\epsilon_{j+1}$, and the purified bits are pushed 
{\em all the way}
to the left. 
This combined $M_j$-and-BCS is repeated $\ell$ times, yielding 
$\langle {\cal L}_{j+1}^{\ell}\rangle  = 
\ell \frac{1+ \epsilon_j^2}{4} \  m$ bits purified
to $\epsilon_{j+1}$. 
For $\ell \ge 4$ we are promised that 
$\langle {\cal L}_{j+1}^{\ell}\rangle > m$, and a CUT operation, 
${\cal C}_{j+1}$, defines
the first $m$ bits to be the output of $M_{j+1}$.

The total number of bits used in $M_{j+1}$ is 
$N_{j+1} = (\ell - 1)m/2 + N_j$ bits, 
where the $N_j$ bits are the ones used at the last
$M_j$ step, and the $(\ell-1)m/2$ bits are the ones previously kept. 
The output of $M_{j+1}$ is defined as the first $m$ bits, and 
in case $M_{j+2}$ is to be performed, 
these $m$ bits are its input.
Let the total number of operations applied at the $j^{\mathrm{th}}$ purification
step, $M_j$, be represented as $T_j$. 
Note that $T_0 = 1$, meaning that $m$ bits are SWAPped with RRTR 
bits in parallel.
Each application of the BCS has a time
complexity smaller than $m^2$ for a near-neighbor connected 
model~(\ref{time-steps-BCS}).
When the $k^{\rm th}$ cooling is done (with $ k \in \{1, \ldots, \ell\}$)
the number of additional steps required to (control-)SWAP 
the adjusted bit at the top of the array
is less than $2(k-1)m$.
Thus we get
$ T_{j+1}  <  \sum_{k=1}^{\ell} [(2\{k-1\}m + 2m)(m/2) +T_j] $.   
Hence, for all $j$, 
\begin{equation}
T_{j+1}  <  \sum_{k=1}^{\ell} [km^2+T_{j}] = 
  \frac{\ell (\ell + 1)}{2} m^2 + \ell T_{j} .
\end{equation}

The purification steps
$M_1$ and $M_2$ can be obtained by 
following the general description of $M_{j+1}$.
For clarity, $M_1$ is described in Figure~1,
$M_2$ is described 
in Figure~2 in appendix~\ref{M-1andM-2},
and both $M_1$ and $M_2$ are described in words 
in that appendix.
For the entire protocol we choose $j_{\rm final}$, and perform 
$M_{j_f}$ starting with $N_{j_f} \equiv n$ bits, and we end up with $m$ bits.

To emphasize 
the recursive structure of this algorithm we use 
the following notations.
[${\cal B}_{\{(k-1) \rightarrow k \}} $]---the BCS 
procedure purifying from $\epsilon_{k-1}$ to $\epsilon_{k}$
(followed by moving the purified bits to the relevant starting point).  
[$\cal S$]---SWAP $m$ bits with the RRTR. 
[${\cal C}_j$]---CUT, keep the first $m$ bits from the 
starting point of the sub-array of the bits with a bias $\epsilon_j$.
Then, 
$M_0 \equiv {\cal S}$, and 
for $j \in \{1, \ldots, j_f\}$
\begin{equation}
M_j  =  {\cal C}_j\;\;\underbrace{
{\cal B}_{\{(k-1) \rightarrow k \}} M_{j-1} \; \cdots 
  \;\;{\cal B}_{\{(k-1) \rightarrow k \}} M_{j-1} \;\;
{\cal B}_{\{(k-1) \rightarrow k \}} M_{j-1} \;\;
}_{\ell\
              \mathrm{times}} ,
\end{equation}
is the recursive formula describing our algorithm.

A full cooling algorithm is $M_{j_f}$ and it is performed starting at
location $\mu=0$.
A pseudo-code for the complete algorithm is shown in Figure~\ref{fig3}.
For any choice of $\epsilon_{\rm des}$, 
one can calculate the required (minimal) $j_f$ such that
$\epsilon_{j_f} \ge \epsilon_{\rm des}$, 
and then $m$ bits (cooled as desired) are obtained
by calling the 
{\bf procedure} COOLING\;$(j_{\mathrm{final}},1,\ell,m)$, 
where $\ell\geq 4$.
We actually use $\ell \ge 5$ in the rest of the paper (although $\ell = 4$ 
is sufficient when the block's size $m$ is very large)
in order to make 
sure that the probability of a successful process does not become too small.
[The analysis done in~\cite{shul_vaz} considers the case in which $m$
goes to infinity, but the analysis does not consider the 
probability of success of the purification in the case where 
$m$ does not go asymptotically to infinity; 
However, in order to motivate experiments in this direction, one 
must consider finite, 
and not too large blocks,
with a size that shall potentially be 
accessible to experimentalists in the near future. In our algorithm, the
case of $\ell = 4$ does not provide a reasonable probability of success for 
the cooling process, but $\ell = 5$ does].

\subsection{Algorithmic Complexity and Error Bound}\label{complex}

\subsubsection{Time and Space Complexity of the Algorithm}

We now calculate $N_f = n$, the number of bits we must start 
with in order to get
$m$ purified bits with bias $\epsilon_{j_f}$. 
We have seen that 
$N_0 = m$ and $ N_j  =  \frac{\ell -1}{2}m+N_{j-1}$,
leading to    
$ N_j=\left(\frac{\ell - 1}{2}j+1\right)m$, and in particular
\begin{equation} 
N_{j_f}=\left(\frac{\ell - 1}{2}j_{\rm final}+1\right)m \ .
\end{equation} 
Thus, to obtain $m$ bits we start with $n= cm$ bits where 
$  c= \frac{\ell - 1}{2}j_{\rm final}+1 $
is a constant depending on the purity we wish to achieve (that is, 
on $j_{\rm final}$)
and on the probability 
of success we wish to achieve (that is, on $\ell$).
For reasonable choices, $j_f$ in the range $3-7$ and $\ell$ in the range $5-7$,
we see that $c$
is in the range $7-22$.
To compare with the Shannon's bound, where the constant goes as
$1/{\epsilon_0}^2$, one can show that here $c$ is a function 
of $1/\log \epsilon_0$.

As we have seen in Section~\ref{cool-algo},
the total number of operations applied at the $j^{\mathrm{th}}$ purification
step, $M_j$, satisfies
$T_j  <  \frac{\ell (\ell + 1)}{2} m^2 + \ell T_{j-1} $.   
Writing $d =  m^2 [\ell (\ell + 1)]/2  $,   
the recursive formula leads to 
$T_{j_f} < \ell^{j_f} T_0 + d  \sum_{j=0}^{j_f-1} \ell^k = 
 \ell^{j_f} + d  [\ell^{j_f} - 1]/[\ell - 1] $.
After some manipulations we get
\begin{equation}
T_{j_f} < m^2  \ell^{j_f + 1} .
\end{equation}
This bound is not tight and a tighter bound can be obtained.
It is also important to mention that in a standard gate-array model
(and even in a ``qubits in a cavity'' model),
in which SWAPs are given almost for free, an order of $m$ instead of $m^2$ is obtained.

Let the relaxation time ${\cal T}_1 $
of the computation bits
be called $ {\cal T}_{\rm comput-bits}$, 
and the relaxation time ${\cal T}_1$
for the RRTR bits be called 
$ {\cal T}_{\rm RRTR}$.
Note that the dephasing time,
${\cal T}_2$, of the computation bits
is irrelevant for our algorithm, and plays a role only {\em after} the cooling
is done.

With the short-term goal in mind we see that $m=20$ can be achieved
(for $\ell = 5$) with $\epsilon_0 = 0.01$, $j_f = 6$, $T_{j_f} < 3.1 \times 10^7$ steps,
and $n = 260$ bits, or with $\epsilon_0 = 0.1$, $j_f = 3$, $ T_{j_f} < 250,000$ steps,
and $n = 140$ bits.
Increasing $m$ to 50 only multiplies the initial length by $2.5$,
and multiplies the time steps by $6.25$.  Thus,  
this more interesting goal can be achieved 
with $\epsilon_0 = 0.01$, $j_f = 6$, $T_{j_f} < 1.9 \times 10^8$ steps,
and $n = 650$ bits, or with $\epsilon_0 = 0.1$, $j_f = 3$, 
$ T_{j_f} < 1.56 \times 10^{6}$ steps,
and $n = 350$ bits.

Concentrating on the case of $j_f = 3$ and $\epsilon_0 = 0.1$,
let us calculate explicitly the timing demands.
For $m=20$ bits, we see that the switching time ${\cal T}_{\rm switch}$
must satisfy $250,000 \  {\cal T}_{\rm switch} \ll {\cal T}_{\rm comput-bits}$ 
in order to allow completion of the 
purification before the system spontaneously relaxes.
Then, with $m^2 = 400$ time steps for each BCS operation,  
the relaxation time for the RRTR bits must satisfy  
$ {\cal T}_{\rm RRTR} \ll 400 \ {\cal T}_{\rm switch}$,
if we want the RRTR bits to be ready when we need them the next time.
As result, a ratio of ${\cal T}_{\rm comput-bits} \gg 625 \ {\cal T}_{\rm RRTR}$
is required in that case.
The more interesting case of $m=50$ demands 
$1.56 \times 10^{6} \ {\cal T}_{\rm switch} \ll {\cal T}_{\rm comput-bits}$, 
$ {\cal T}_{\rm RRTR} \ll 2500 \ {\cal T}_{\rm switch}$,
and ${\cal T}_{\rm comput-bits} \gg 625 \ {\cal T}_{\rm RRTR}$.
Note that choosing $\ell=6$ increases the size by a factor of $5/4$, and the time
by a factor of $6^4 / 5^4 \approx 2$.
We shall discuss the possibility of obtaining these numbers in an actual experiment
in the next section.

\subsubsection{Estimation of error}

Since the cooling algorithm is probabilistic, and so far we have considered only the
{\em expected} number of purified bits, we need to make sure that in practice
the actual number of bits obtained is larger than $m$ with a high probability. 
This is especially important when one wants to deal with
relatively small numbers of bits. 
We recall that the random variable ${\cal L}^k_j$ is the number of 
bits purified to $\epsilon_j$, 
after the $k^{\mathrm{th}}$ round of purification step--$M_{j-1}$ 
each followed by 
${\cal B}_{\{(j-1)\rightarrow j \}}$. Hence, prior to the CUT ${\cal C}_j$
we have  ${\cal L}_j^\ell$ bits with bias $\epsilon_j$, where the
expected value $\langle {\cal L}_j^\ell \rangle =
 \ell \frac{1+ \epsilon_{j-1}^2}{4} \  m > \frac{ \ell m}{4}$, and
we use $\ell\geq 5$. 
Out of these bits we keep only the first $m$ qubits, i.e., we keep
at most a fraction $\frac{4}{\ell}$ of the average length of the string of
desired qubits.  Recall also that ${\cal L}_j^\ell$ is a sum of independent
Bernoulli random variables, and hence one can apply a suitable form of the
strong law of large numbers to determine the probability of success, i.e., the
probability that ${\cal L}_j^\ell \geq m $.

The details of applying a law of large numbers are given in 
appendix~\ref{error-probability}.
Here we only state the result.
Chernoff's bound implies that
the probability of {\em failing} to get at 
least $m$ bits with bias $\epsilon_j$ is
\[ \Pr\left[{\cal L}_j^\ell < m\right] \le \exp\left(-\frac{1}{2} \left(1-
\frac{4}{\ell}\right)^2 \frac{\ell}{4} m\right)= \exp\left(-
\frac{(\ell-4)^2}{8\ell } m\right). \]
For the probability of success of the entire algorithm we have  
the following conservative lower bound
\begin{equation}
\Pr\left[\mbox{success of the algorithm}\right]   \ge
\left[1-\exp\left(-\frac{(\ell-4)^2}{8\ell}m\right)\right]^{
(\ell^{j_f}-1)/(\ell-1) } .  
\end{equation}

The probability of success is given here for several interesting cases with   
$j_f = 3$ (and remember that the probability of success increases 
when $m$ is increased):
For $m=50$ and $\ell = 6$ we get  
$\Pr\left[\mbox{success of the algorithm}\right]   > 0.51$.
For $m=50$ and $\ell = 5$ we get  
$\Pr\left[\mbox{success of the algorithm}\right]   > 2.85 \times 10^{-5}$.
This case is of most interest due to the reasonable time scales.
Therefore, it is important to mention here
that our bound is very conservative since we demanded success in {\em all}
truncations (see details in appendix~\ref{error-probability}), 
and this is not really required in practice. 
For instance, if only $m-1$ bits are purified to $\epsilon_1$ in one round
of purification,
but $m$ bits are purified in the other $\ell - 1$ rounds, then
the probability of having $m$ bits at the resulting $M_2$ process is not zero,
but actually very high.
Thus, our lower bound presented above should not discourage 
the belief in the success of this algorithm, since a much 
higher probability of success is actually expected.

\section{Physical Systems for Implementation}

\label{sec-exp}

The spin-refrigeration algorithm relies on the ability to combine rapidly relaxing 
qubits and slowly relaxing qubits in a single system. T1 lifetimes of atomic 
spins in molecules can vary greatly, depending on the degree of isolation from 
their local environment. 
Nuclei that are positioned close to unpaired electrons, for example, can 
couple strongly to the spin of these electrons and decay quickly. Identical 
nuclei that are far removed from such an environment can have extremely long 
lifetimes. Many examples of T1 varying over three orders of magnitude, from seconds 
to milliseconds, exist in the literature. One example is the $^{13}$C nuclear 
relaxation rate, which changes by three orders of magnitude depending
on whether the $^{13}$C atom is part of a phenoxyl or triphenylmethyl 
radical~\cite{kurreck}. By combining these different chemical 
environments in one single molecule, and furthermore, by making use of
different types of nuclei, one could hope to achieve even a ratio of $10^4$.

Another possible choice is to combine the use of nuclear 
spins and electron spins. The coupling between nuclei and 
electrons that is needed to perform the desired SWAP operations, has been well 
studied in many systems by the Electron-Nuclear Double Resonance (ENDOR) 
technique\cite{kurreck}. The electron spins which typically interact strongly with 
the environment could function as the short lived qubits, and the nuclei as the 
qubits that are to be used for the computation. In fact, the relaxation rate of electrons
is commonly three orders of magnitude faster than the relaxation rate of nuclei in the
same system, and another order of magnitude seems to be easy to obtain.
Note also that the more advanced TRIPLE resonance technique 
can yield significantly better results than the ENDOR technique\cite{mobius}.

This second choice of strategy has another advantage---it allows 
initiation of the the process
by SWAPing the electron spins with the nuclear spins, thus getting 
much closer to 
achieving the desired initial bias,
$\epsilon_0 = 0.1$, which is vital for allowing reasonable time-scales
for the process.
If one achieves ${\cal T}_{\rm comput-bits} \approx 10 {\rm secs}$,
and ${\cal T}_{\rm RRTR}  \approx 1 {\rm millisecs}$,
then a switching time of $\approx 10 {\rm microsecs}$ allows our 
algorithm to yield 20-qubit computers. 
If one achieves ${\cal T}_{\rm comput-bits} \approx 100 {\rm secs}$,
and ${\cal T}_{\rm RRTR}  \approx 10 {\rm millisecs}$,
then a switching time of $\approx 10 {\rm microsecs}$ allows our 
algorithm to yield 50-qubit computers.

\section{Discussion} \label{conc}

In this paper we suggested 
``algorithmic cooling  via polarization heat bath'' 
which removes entropy into the environment, and allows compression beyond the Shannon's bound.
The algorithmic cooling can solve the scaling problem of NMR quantum computers,
and can also be used to refrigerate spins to very low temperatures.
We explicitly showed how, using SWAP operations between electron spins and nuclear spins,
one can obtain a 50-qubit NMR quantum computer, starting with 350 qubits, and using
feasible time scales.
Interestingly, the interaction with the environment, usually a most undesired interaction,
is used here to our benefit.

Some open questions which are left for further research:
(i)~Are there better and simpler cooling algorithms?
(ii)~Can the above process be performed in a (classical) fault-tolerant way?
(iii)~Can the process be much improved by using more sophisticated compression algorithms?
(iv)~Can the process be combined with a process which resolves the addressing problem?
(v)~Can one achieve sufficiently different thermal relaxation times for the two 
different spin systems? 
(vi)~Can the electron-nuclear spin SWAPs be implemented on the same systems
which are used for quantum computing?
Finally, the summarizing question is: (vii)~How far are we from demonstrating
experimental algorithmic cooling, and how far are we from using it to yield
20-qubit, 30-qubit, or even 50-qubit quantum computing devices?

\pagebreak

\includegraphics{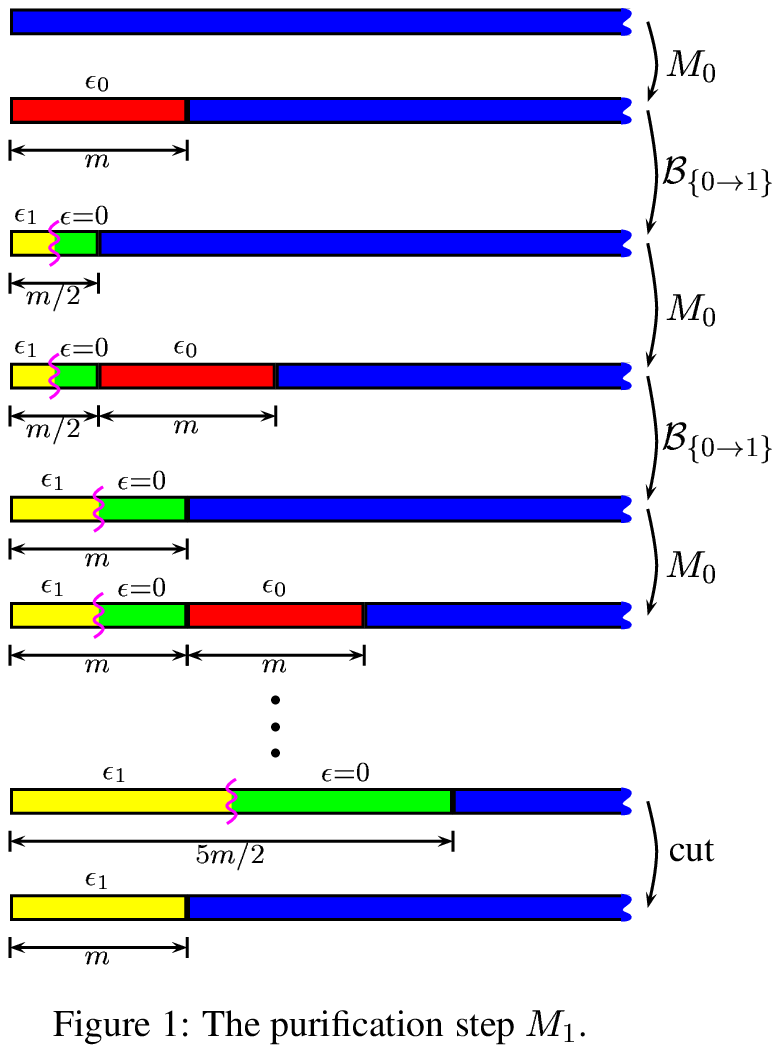}

\includegraphics{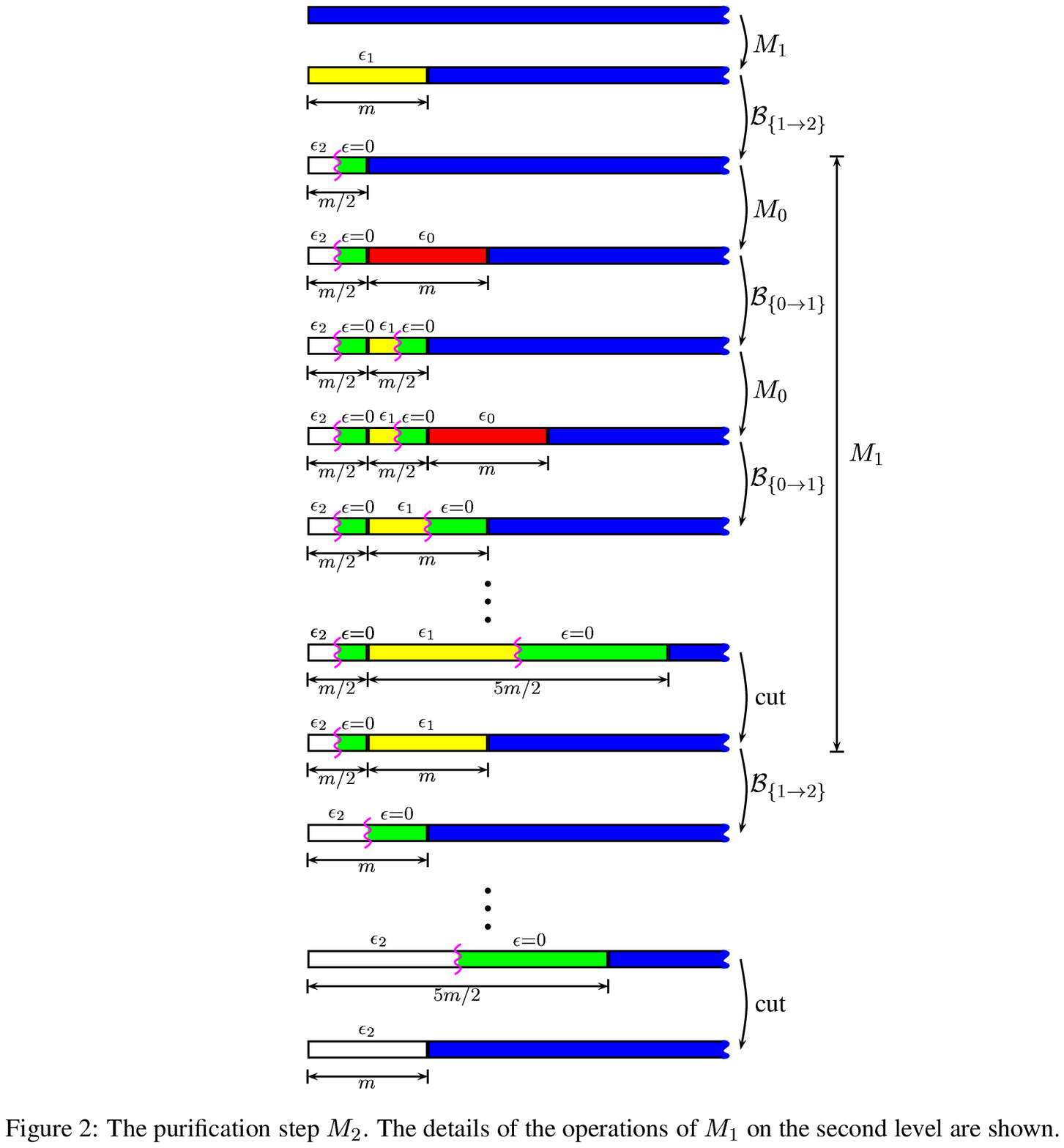}

\setcounter{figure}{2}

{\small
\begin{figure}\label{fig3}

\begin{center}

\fbox{\parbox[c]{0.95\linewidth}{
\underline{{\bf procedure}
 $\mbox{COOLING}\,(j,\mu,\ell,m)$
}

\hspace{16mm}\parbox{0.85\linewidth}{\begin{itemize}

\item[({\bf comment:}] this procedure returns $m$ bits with bias $\epsilon_j$
starting from the bit in position $\mu$; $\ell$ is the cooling depth)

\end{itemize}}

{\bf begin}
 \hspace{4mm} {\bf if} $j=0$ {\bf then do}

      \hspace{16mm} {\bf call} $\mbox{SWAP}\,\left(
       \mu, m \right) $

 \hspace{4mm} {\bf else do}

   \hspace{8mm} {\bf begin}

     \hspace{12mm} {\bf for} $\mbox{\tt depth}=0$ {\bf to} $\ell-1$ {\bf do}

     \hspace{16mm} {\bf begin}

     \hspace{20mm} {\bf call} $\mbox{COOLING}\,
     \left(j-1,\mu+(\mbox{\tt depth})*\frac{1}{2}m,\ell,m\right)$

     \hspace{20mm} {\bf call} $\mbox{BCS}\,\left(\mu+\mbox{\tt
           depth}*\frac{1}{2}m, \mu \right)$

     \hspace{16mm} {\bf end}

   \hspace{8mm} {\bf end}

{\bf end}

\vspace{2.5mm}
\underline{{\bf procedure} $\mbox{BCS}\,(\nu,\nu_0)$
}

{\bf begin}

\parbox{0.9\linewidth}{\begin{itemize}
\item[ ] Apply the BCS to the $m$ bits starting at location $\nu$, and push the
purified bits always to the location $\nu_0$ (where, $\nu_0\leq \nu$).
\end{itemize}}

{\bf end}

\vspace{2.5mm}
\underline{{\bf procedure} $\mbox{SWAP}\,(\mu,m)$
}

{\bf begin}

\parbox{0.9\linewidth}{\begin{itemize}
\item[ ] Perform a cooling operation by swapping the bits at location
$\mu$ to $\mu+m-1$ with the RRTR bits and thus resetting their bias to
$\epsilon_0$.
\end{itemize}}

{\bf end}
}}
\caption{A pseudo--code for the cooling algorithm.}
\end{center}
\end{figure}
  }

\pagebreak

\mbox{\vspace{1cm}}

\begin{table}
\begin{center}
\begin{tabular}{|c|c|c|c|c|c|c|} \hline
 & & & & \multicolumn{3}{c|}{ $p\approx (1-\delta_f)^m$ }\\ \cline{5-7}
 \raisebox{1.6ex}[0cm][0cm]{$\epsilon_0$} & 
 \raisebox{1.6ex}[0cm][0cm]{$j_f$} & 
 \raisebox{1.6ex}[0cm][0cm]{$\epsilon_f$} 
 & \raisebox{1.6ex}[0cm][0cm]{$\delta_f=\frac{1-\epsilon_f}{2}$} 
 & $m=20$ & $m=50$  & $m=200$ \\ \hline\hline
                                 & 0 & 0.1        & 0.45     
& 6.4 $\times 10^{-6}$ & unfeasible           & unfeasible       \\ \cline{2-7}                  0.1            & 3 & 0.666      & 0.1672   
& 2.6 $\times 10^{-2}$ & 1.1 $\times 10^{-4}$ & unfeasible       \\ \cline{2-7}                                 & 4 & 0.922      & 0.0388   
& 4.5 $\times 10^{-1}$ & 1.3 $\times 10^{-1}$ & 3.7 $\times 10^{-4}$  \\ \hline
                                 & 0 & 0.01       & 0.495    
& 1.2 $\times 10^{-6}$ & unfeasible           & unfeasible       \\ \cline{2-7}
                  0.01           & 6 & 0.565      & 0.2175   
& 7.4 $\times 10^{-3}$ & 4.7 $\times 10^{-6}$ & unfeasible       \\ \cline{2-7}
                                 & 7 & 0.856      & 0.0718   
& 2.2 $\times 10^{-1}$ & 2.4 $\times 10^{-2}$ & 3.4 $\times 10^{-7}$  \\ \hline     
\end{tabular}
\end{center}
\caption{Feasibility of running an $n$-qubit NMR computer, when 
the polarization bias is improved to $\epsilon_f$, 
prior to using the PPS technique.}
\end{table}

\pagebreak

\appendix

\section{PPS technique and the scaling problem} \label{PPS-tech}

To illustrate the PPS scheme, let us first consider the case of $n=1$ qubits,
with an arbitrary bias $\epsilon$.
The initial state is given by equation (\ref{initial-state})
[with $\epsilon$ replacing $\epsilon_0$],
\begin{equation} \label{init-1b-state}
  \rho_{\epsilon} = 
  \begin{pmatrix}
               (1+\epsilon)/2 & 0 \\
               0                & (1-\epsilon)/2
  \end{pmatrix} 
= 
  \begin{pmatrix}
               \epsilon    & 0 \\
               0           & 0
  \end{pmatrix}  + 
 \begin{pmatrix}
              (1-\epsilon)/2 & 0 \\
              0                & (1-\epsilon)/2
  \end{pmatrix} , 
\end{equation}
where the second form is already in the form of a PPS,
$ \epsilon |0\rangle\langle0| + [(1 - \epsilon)/2] {\cal I} $.

Let us now consider the case of $n=2$ qubits.
Since the initial state of each qubit is given by equation (\ref{init-1b-state}),
the density matrix of the initial thermal-equilibrium state of the two-qubit
system can be represented as:
\begin{equation} \label{init-2b-state-a}
\rho_{\mathrm{init}}^{n=2}=
  \begin{pmatrix}
     (1+\epsilon)/2 & 0 \\
     0                & (1-\epsilon)/2
  \end{pmatrix}
 \otimes
  \begin{pmatrix}
    (1+\epsilon)/2 & 0 \\
    0                & (1-\epsilon)/2
  \end{pmatrix} .
\end{equation}
For the purpose of understanding the PPS it is legitimate to ignore the difference 
between the $\epsilon$ of the two spins, but in practice they must differ a bit, 
since the only way to address one of them and not the other is by using accurate 
fields such that only one level splitting is on resonance with that field.

For the purposes of generating
the PPS, it is instructive to represent the initial
state as
\begin{equation} \label{init-2b-state-b}
  \rho_{\mathrm{init}}^{n=2}=\frac{(1+\epsilon)^2}{4} \rho_{00} +
  \frac{1-{\epsilon}^2}{4} \rho_{01} +
  \frac{1-{\epsilon}^2}{4} \rho_{10} +
  \frac{(1-\epsilon)^2}{4} \rho_{11} ,
\end{equation}
where $\rho_i=\vert i\rangle\langle i\vert$, and $i$ is being a binary string.
The coefficient of each $\rho_i$, say $P_i$, is the
probability of obtaining the string $i$ in a measurement (in the computation basis)
of the two qubits. 
Thus, $P_{00} = (1 + \epsilon)^2/4$, 
$P_{01} = P_{01} = (1 - \epsilon^2)/4$, 
and $P_{11} = (1 - \epsilon)^2/4$. 
In order to generate a pseudo-pure state, let us
perform one of the following three transformations:
$S_1 =  I$ (the identity),
\[
S_2=\left \{ \begin{array}{ccc}
     |00\rangle & \longrightarrow & |00\rangle \\
     |01\rangle & \longrightarrow & |10\rangle \\
     |10\rangle & \longrightarrow & |11\rangle \\
     |11\rangle & \longrightarrow & |01\rangle
    \end{array} \right . \qquad
S_3=\left \{ \begin{array}{ccc}
     |00\rangle & \longrightarrow & |00\rangle \\
     |01\rangle & \longrightarrow & |11\rangle \\
     |10\rangle & \longrightarrow & |01\rangle \\
     |11\rangle & \longrightarrow & |10\rangle
    \end{array} \right .
 \]
with equal probability, so that each molecule (each computer)
is subjected to one of the above-mentioned
transformations. This transformation can be carried out in an experiment by
applying different laser pulses to different portions of the liquid, or by
splitting the liquid into three portions, applying one of the transformations to
each part, and mixing the parts together. These permutations map  $|00\rangle$
to itself, and completely mix all the other states. As a result,
the density matrix of the final state becomes:
\begin{eqnarray*}
\rho_{\mathrm{pps}}^2 & = & \frac{(1+\epsilon)^2}{2^2} \rho_{00} +
\frac{1-(1+\epsilon)^2/2^2}{2^2-1}(\rho_{01}+\rho_{10}+\rho_{11})\\
 &=& \frac{ (1+\epsilon)^2 - 1}{2^2-1} \rho_{00} +
\frac{1-(1+\epsilon)^2/2^2}{2^2-1}(\rho_{00}+\rho_{01}+\rho_{10}+\rho_{11})\\
 &=& \frac{ (1+\epsilon)^2 - 1}{2^2-1} \vert 00\rangle\langle 00\vert +
 \frac{2^2-(1+\epsilon)^2}{2^2-1} \; {\cal I}\ ,
\end{eqnarray*}
so that finally, $p= \frac{ (1+\epsilon)^2 - 1}{2^2-1}$ is the probability
of having the pure state. Note that $p$ is not $P_{00}$ since the
completely mixed state also contains a contribution from $P_{00}$.

The above procedure for mixing can be directly generalized to a system
comprising $n$ qubits\footnote{In practice, one might accomplish an 
approximate mixing instead, 
due to the exponential number of different rotations required for a perfect
mixing.}.  The density matrix for the final state is:
\begin{eqnarray*}
\rho_{\mathrm{pps}}^n & = & \frac{(1+\epsilon_0)^n}{2^n} \rho_{00\ldots 0} +
\frac{1-(1+\epsilon_0)^n/2^n}{2^n-1} \left(\sum_{i=1}^{2^n-1}
\rho_{i}\right)\\
& = & \frac{(1+\epsilon_0)^n-1}{2^n-1} \vert 00\cdots0\rangle\langle 00\cdots0\vert +
\frac{2^n-(1+\epsilon_0)^n}{2^n-1} {\cal I}\ . \end{eqnarray*}
The probability of the $n$ bit pure state  
$|00\cdots 00\rangle$ is $p = 
\displaystyle \frac{(1+\epsilon_0)^n-1}{2^n-1}$
which is $p \approx (n \epsilon_0)/2^n$ for small $\epsilon$,
hence exponentially small with $n$.
Obviously, such a signal is highly
obscured by the completely mixed state, leading to an exponentially small
signal-to-noise ratio, and hence, to the scaling problem.
However, it is clear now that information is lost in the process, due to the mixing step:
In order to obtain the PPS we need to ``forget'' the transformation done on each 
computer, and consider only the average result.

In order to clearly see
the inherent loss of information in the mixing process, consider an ensemble
computer  in its initial state, $\rho_{\mathrm{init}}^n$. We note that one can
perform any single qubit operation (and measure)  on any of the $n$ qubits {\em
without} any purification.  For example, if one were to measure any individual
qubit in the ensemble when it is in its initial state, then one would observe a
$|0\rangle$,  
irrespective of how large $n$ is;
similarly, one can perform single qubit rotations and then make measurements
without any purification of the initial state. The same is not true if the
rotation is applied to the 
PPS  $\rho_{\mathrm{pps}}^n$; then the
completely mixed state dominates, and the  exponentially small
signal is obscured. 
Unfortunately, performing 2-qubit computation (or more) with mixed states is not 
a realistic choice.
To summarize, the PPS technique {\em causes} the problem of scaling, by
losing information {\em on purpose}.

\section{A detailed description of $M_1$ and $M_2$} \label{M-1andM-2}

In the purification step $M_1$ we wish to obtain $m$ bits with a bias $\epsilon_1$.
In order to achieve this
we apply $\ell$ cooling operations (SWAPs with RRTR), each followed by repeated
applications of the BCS (acting on bits with $\epsilon_0$ bias). 
This is done as follows:
The $m$ bits at the head of the array (positions 1 to $m$) are SWAPped with RRTR to
yield $\epsilon_0$. Then, a BCS is applied onto them, resulting in 
having ${\cal L}_1$ purified bits at the left,
unpurified adjusted bits next to them, and finally, the supervisor bits at the positions
$m/2+1 $ to $m$ (the  
right locations of the $m$-bit-array).
Then a similar set of operations is applied to an array of $m$ bits at locations
$m/2 +1$ to $3m/2$. This array includes all the supervisor bits of the previous operation
plus $m/2$ more bits. First, these $m$ bits are reset to $\epsilon_0$. 
When a BCS is applied onto these $m$ bits, all purified bits are 
pushed to the left, but now it is a push to the head of the entire $\epsilon_1$-bias 
string, 
all unpurified adjusted bits are kept in their place and all supervisor bits are 
pushed to the right of the $m$-bit array. 
Pushing the purified bits {\em all the way} to the left is vital, since we want
to be certain that no unpurified bit remains among the purified bits.
Let us denote the number of purified bits at the end of this step
${\cal L}_1^2$ where the superscript is added to indicate it is a count done after
a second SWAP with RRTR. [Thus, the number of bits after the first SWAP with RRTR
is renamed ${\cal L}_1^1$.] The same set of operations is repeated $\ell$ times,
and at its end the entire array used for $M_1$ 
contains $ N_1 = (\ell-1) m /2 + m$ bits, where the $m$ bits are the ones used at 
the last compression, and the $(\ell-1)m/2$ bits are the ones previously kept. 
Of these $N_1$ bits,  
$\langle {\cal L}_1^{\ell}\rangle $ purified bits are at the left, and $m/2$ dirty
supervisor bits are at the right (remaining from the 
last application of BCS).
The expectation value for the length of the purified bits satisfies
$\langle {\cal L}_1^{\ell}\rangle  = \ell \frac{1+ \epsilon_0^2}{4} \  m$.
Finally, we define the output of this purification step to be the first $m$ bits at the left.
Then, for $\ell \ge 4$, 
$\langle {\cal L}_1^{\ell}\rangle  > m$.

In the purification step $M_2$ we wish to obtain $m$ bits with a bias $\epsilon_2$.
In order to achieve this goal,
we apply $\ell$ purification steps $M_1$, each followed by 
a BCS applied to exactly $m$ bits at a bias $\epsilon_1$:
First, $M_1$ is applied onto $N_1$ bits yielding an output of $m$ bits at
a bias $\epsilon_1$, then a BCS is applied onto these bits yielding
a string of expected length $\langle {\cal L}_2^1 \rangle = \frac{1+ \epsilon_1^2}{4} \  m$
bits purified to a bias $\epsilon_2$ and pushed to the left.
Then $M_1$ is applied again to an array of $N_1$ bits, starting at the location
$m/2 + 1$.  This time all BCS operations {\em within} $M_1$ push the bits to the first
location of that array (the location $m/2 +1$ of the entire string).
At the end of the second $M_1$ application, a BCS is applied to $m$ bits at a bias
$\epsilon_1$ purifying them to $\epsilon_2$, and the purified bits are pushed all the way
to the left. 
Then $M_1$ is applied a third time to an array of $N_1$ bits, starting at location
$m + 1$.  This time all BCS operations {\em within} $M_1$ push the bits to the first
location of the array (the location $m +1$ of the entire string).
At the end of the third $M_1$ application, a BCS is applied to $m$ bits at a bias
$\epsilon_1$ purifying them to $\epsilon_2$, and the purified bits are pushed all the way
to the left. 
This combined $M_1$-and-BCS is repeated $\ell$ times, yielding 
$\langle {\cal L}_2^{\ell}\rangle  = \ell \frac{1+ \epsilon_1^2}{4} \  m$ bits purified
to $\epsilon_2$. 
The total number of bits used in $M_2$ is 
$N_2 = (\ell - 1)m/2 + N_1$ bits, 
where the $N_1$ bits are the ones used at the last
$M_1$ step, and the $(\ell-1)m/2$ bits are the ones previously kept. 
The output of $M_2$ is defined as the first $m$ bits.

\section{Probability of error in the algorithm} \label{error-probability}

We utilize the following form of the {\em Chernoff's
bound}: If $X_1,\ldots,X_t$
are $t$ independent random variables with $\Pr({X_i=1})=p$ and $\Pr({X_i=0})=1-p$,
then for $X=X_1+\cdots+X_t$ we have
\begin{equation}
  \Pr[{X<tp(1-a)}] < \exp\left(-\frac{a^2 tp}{2}\right).
\label{chernoff}
\end{equation}
In our case, the number of trials $t$, is $\ell m$.  The probability of
keeping a bit (so that $X_i=1$) is $p=(1+{\epsilon_{j-1}}^2)/4$ which is greater than $1/4$.
We set $a=1-\frac{1}{\ell p}$ so that $t p (1-a) = m$.
Therefore, $a>1-\frac{4}{\ell}$. Now, using the 
fact that $p>1/4$, Chernoff's bound (\ref{chernoff}) implies that
the probability to {\em fail} to get at 
least $m$ bits with bias $\epsilon_j$ is
\[ \Pr\left[{\cal L}_j^\ell < m\right] < \exp\left(-\frac{1}{2} \left(1-
\frac{4}{\ell}\right)^2 \frac{\ell}{4} m\right)= \exp\left(-
\frac{(\ell-4)^2}{8\ell } m\right). \]

In the complete algorithm that runs for $j_{\mathrm{final}}$ purification
steps, we need to calculate the total number of times the above--mentioned hard
truncations are performed and demand success in all\footnote{This is a very 
conservative demand, so actually the probability of success is much higher than the 
one we calculate here.} of them. In other words, to get ${\cal L}_j^\ell$ purified
bits at purification step $M_j$ (from which $m$ bits will be taken via another truncation) 
we first need to successfully provide $\ell$ times $m$--bit strings with bias
$\epsilon_{j-1}$. The recursive nature of our algorithm demands the successful 
purification of all $m$--bit 
strings with smaller biases $\epsilon_k$, for all $0<k<j$, in order to achieve this goal
for the $M_j$ step.
Let $C_j$ be the number of all $m$--bit strings with biases smaller than 
$\epsilon_j$, needed
at the $j^{\mathrm{th}}$ step. Recall that the $m$ bits at $\epsilon_0$ are given with 
certainty, so only one successful truncation is required to get $\epsilon_1$.
Then
\begin{eqnarray*}
  C_1 & = & 1, \\
  C_j & = & 1+\ell C_{j-1}.
\end{eqnarray*}
Hence,
$  C_j= \sum_{k=0}^{j-1} \ell^k = \frac{\ell^j-1}{\ell-1} $,  and
\begin{equation}
 C_{j_f}
= \frac{\ell^{j_f}-1}{\ell-1} \ . 
\end{equation}
For the probability of success of the entire algorithm we demand success
in all the $C_{j_{\mathrm f}}$ truncation processes
\begin{eqnarray*}
\Pr\left[\mbox{success of the algorithm}\right]   &>& 
\left( 1-\Pr\left[{\cal L}^\ell_j<m \right]\right)^{C_{j_f}} \\   &>&
\left(1-\exp\left(-\frac{(\ell-4)^2}{8\ell}m\right)\right)^{ 
(\ell^{j_f}-1)/(\ell-1)} .  
\end{eqnarray*}


\begin{thebibliography}{00}
\bibitem{diff-bias}  Individual addressing of
qubits requires a slightly different bias for each one, which is easily 
achievable in practice.
\bibitem{bits-qubits} We refer to these qubits as bits since no quantum 
effects are used in the cooling process.
\bibitem{ell-j}
An algorithm with $\ell$ replaced by $\ell_j$ (different numbers of repetitions,
depending on the bias-level $j$) 
could have some advantages, but will not be as easy to analyze.
\bibitem{cooling} Actually, 
adjusted bits which failed to purify are always dirtier than the 
RRTR bits, but supervisor 
bits are dirtier only as long as $\epsilon_j^2 < \epsilon_0$. 
Therefore the {\em CUT} of the adjusted bits which failed to purify, ${\cal C}$
(which is explained in the next subsection) is the main ``engine''
which cools the NMR system at all stages of the protocol.
\end{thebibliography}
\end{document}